\documentclass[aps,prl,twocolumn]{revtex4}

\usepackage[utf8]{inputenc}

\usepackage{textcomp}
\usepackage{graphicx}
\usepackage{latexsym}
\usepackage{amsfonts}
\usepackage{amsmath}
\setkeys{Gin}{width=0.55\textwidth}

\begin{document}

\title{Experimental observation of  bound states of 2D Dirac electrons on the 
 surface of  topological insulator  Bi$_2$Se$_3$}
\author{N.I.~Fedotov}
\author{S.V.~Zaitsev-Zotov}
\affiliation{Kotel'nikov IRE RAS, Mokhovaya 11, bld.7, 125009 Moscow, Russia}

\date{\today}

\begin{abstract}
Topologically protected surface states of three-dimensional topological 
insulators provide a model 
framework for studying massless Dirac electrons in two dimensions.
Usually a step on the surface of a topological insulator is treated as a scatterer for the Dirac electrons, and the study of its effect is focused on  the interference of the incident and scattered electrons. Then a major role  plays the warping of the Dirac cone far from the Dirac point. Here we experimentally demonstrate the existence  of another significant effect  near the Dirac point brought about by the presence of steps. Namely the band bending in the vicinity of steps leads to formation of 1D bound states in the corresponding potential wells.  
We report the observation of bound states in such potential wells in our 
scanning tunneling microscopy and spectroscopy  investigation of the surface of the topological insulator Bi$_2$Se$_3$. Numerical simulations support our conclusion and
provide a recipe for their identification.
\end{abstract}

\maketitle

Recently, a number of solid state systems have been used as model systems for 
investigating exotic 
particle physics for quantum field theory and high-energy physics. 
Of great interest in this respect are Dirac materials~\cite{Wehling2014}. Their 
electronic excitations obey the Dirac equation, in place of the Schroedinger 
one. This opens up a possibility to study quasi-relativistic physics in a 
convenient tunable solid-state setting.
In particular, graphene and, more recently, topological insulators provide a 
model framework for studying massless Dirac electrons in two dimensions.

Three-dimensional topological insulators are characterized by the presence of 
gapless surface states on the background of an insulating bulk \cite{review}. 
The existence of these states is governed by the nontrivial value of the 
$\mathbb{Z}_{2}$ topological invariant. 
In the prototypical topological insulator Bi$_2$Se$_3$ \cite{Zhang2009} the topologically protected surface 
states form a cone in $k$-space. The  apex of the cone (the Dirac point) is 
located at the $\Gamma$-point of the surface 
Brillouin zone, its energy being within the bulk band gap. 
In the vicinity of the Dirac point the Bi$_2$Se$_3$ surface states can be described by a 
model Dirac Hamiltonian $H=A\boldsymbol{\sigma{}k}$ \cite{Zhang2009}. 
Here $\boldsymbol{\sigma}=(\sigma_x, \sigma_y)$ is the Pauli matrices vector, 
$\boldsymbol{k}$ is the wave vector, $A\approx0.33$~eVnm defines the Fermi 
velocity.

Massless Dirac electrons possess a number of peculiar properties. For instance, 
they can travel without reflection through a potential step (Klein tunneling 
\cite{Klein1929, Katsnelson2006}).
It is generally accepted that due to the Klein tunneling the confinement of 
massless 
fermions by means of purely electrostatic potential is not possible.
It is true in a one-dimensional (1D) case, however for  a 2D Dirac system with 
a 1D potential, states localized in one 
direction (perpendicular to the potential well or barrier) exist 
\cite{Pereira2006, Tudorovskiy2007}.
The issue of Dirac electrons confinement continues to attract considerable 
attention from theorists \cite{Yampolskii2008, 
Yokoyama2010, Seshadri2014, Hartmann2017}. Experiments in this area have been 
mostly concentrated on graphene: from Klein tunneling~\cite{Young2009} to 
lithographically defined quantum dots~\cite{Ponomarenko2008} and chemically 
synthesized 
flakes~\cite{Subramaniam2012}. More  recently, quasi-bound states were observed 
by means of 
scanning tunneling microscopy in  electrostatically defined quantum dots 
\cite{Gutierrez2016, Lee2016}. Current distribution in one-dimensional graphene 
edge channels was investigated in transport measurements \cite{Allen2016}. 

Topologically nontrivial systems offers an extensive playground for studying exotic quasiparticle physics. A prime example of the reach variety of physical properties in these systems are sister compounds Bi$_2$Se$_3$ and Bi$_2$Te$_3$, both topological insulators. The Dirac cone warping and the Dirac point position in the bulk valence band in the latter substance cause a dramatic difference in the behavior of the topologically protected surface states. As shown by tight binding calculations \cite{Kobayashi}, edge states could form on the surface steps of Bi$_2$Te$_3$, whereas no such states are expected on Bi$_2$Se$_3$. These calculations do not, however, take into account
the 1D potential wells for the Dirac electrons that arise on
the surface of the topological insulator Bi$_2$Se$_3$ as band
bending occurs near surface steps \cite{Fedotov2017}.

Steps on the surface of topological insulators are interesting objects of experimental \cite{Alpichshev, Interf1, Interf2, Dmitriev, Fedotov2017, Seo2010, Bauer2016} as well as theoretical \cite{Yokoyama2010, Seshadri2014, Kobayashi, Xu2018, Moon, Narayan2014, Alos2013, Wang2010, Biswas2011, Liu2012, Zheng2012, An2012} 
 investigation.
Edge states were found on steps in crystalline topological insulators \cite{crystalTI} and Weyl  semimetals \cite{Weyl1, Weyl2, Weyl3}.
Topological insulators of higher order revealing hinge states are also at the forefront of topological insulator research \cite{HighTI}.

Most STS studies of surface steps in bismuth chalcogenide systems focus on the surface states scattering  and interference patterns \cite{Interf1, Interf2}.
Alpichshev {\em et al.} \cite{Alpichshev} observed an accumulation of LDOS near a step on the surface of topological insulator Bi$_2$Te$_3$ by means of STS. This accumulation was ultimately attributed to a contribution from the states on the side surface of the step \cite{AlpichshevDiss}.

Formation of waveguide states on the side surface of a step in Bi$_2$Se$_3$ was discussed theoretically in \cite{Moon} using an optical analogy on the basis of the lower Fermi velocity of the topologically protected surface states on the side surface obtained in DFT calculations.

Other theoretical approaches include treating the step as a $\delta$-function potential barrier \cite{Biswas2011, Liu2012, Zheng2012, An2012, Xu2018}. 
A barrier potential at the step  results in bound states formation, but branches of  $E(k_y)$ point in the opposite direction compared to the states in a potential well discussed here.
Papers \cite{Biswas2011, Liu2012, Zheng2012, An2012} focus on the scattering of the surface states rather than the formation of bound states. 

Here we report  direct observation  of bound states in  potential 
wells formed in the vicinity of steps on the surface of the topological insulator Bi$_2$Se$_3$.
We employ scanning tunneling microscopy and spectroscopy  
(STM/STS) to image the spatial distribution of LDOS and numerical modeling to support our findings.

 \begin{figure}[h]
 \includegraphics[width=0.2\textwidth]{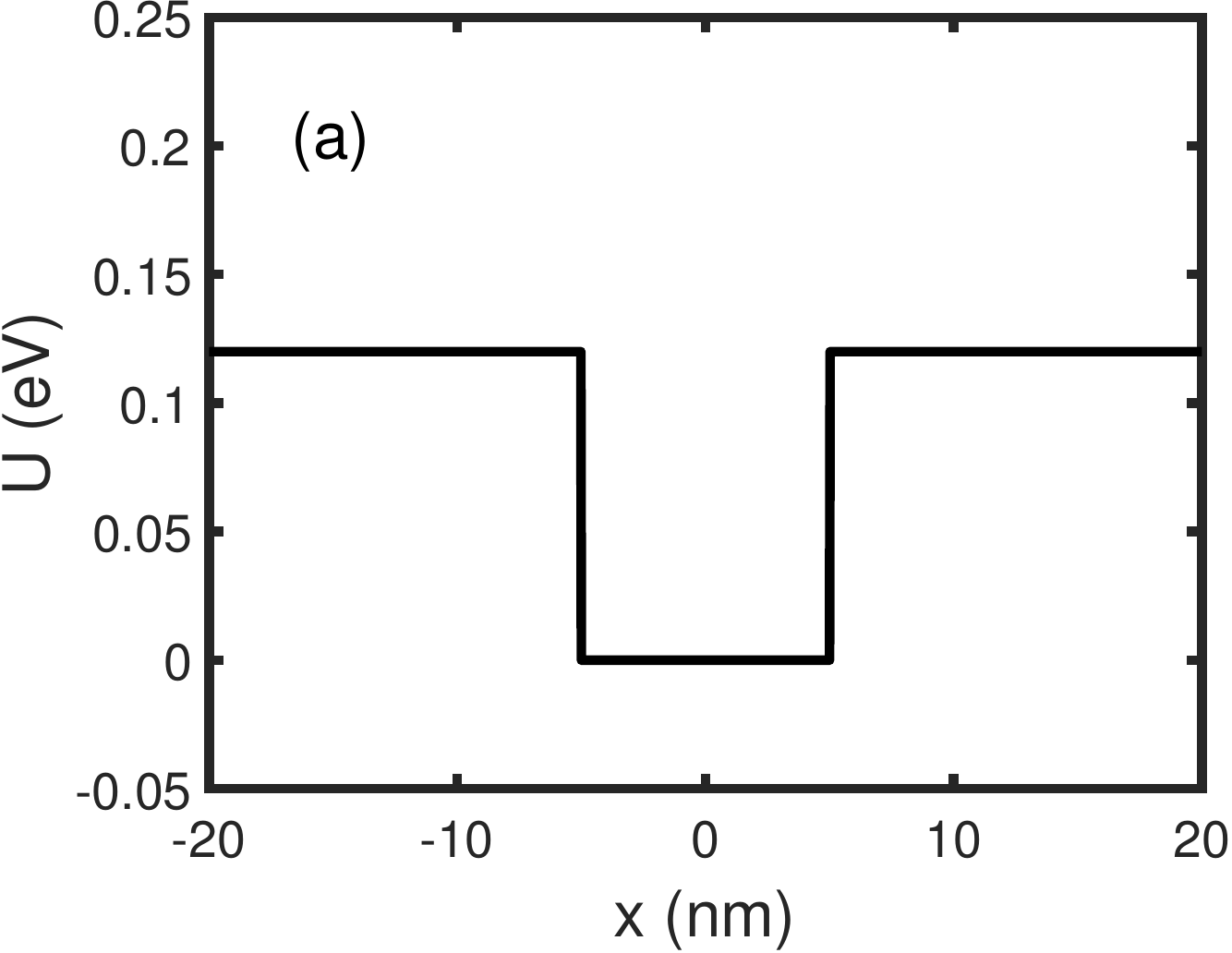}
 \includegraphics[width=0.2\textwidth]{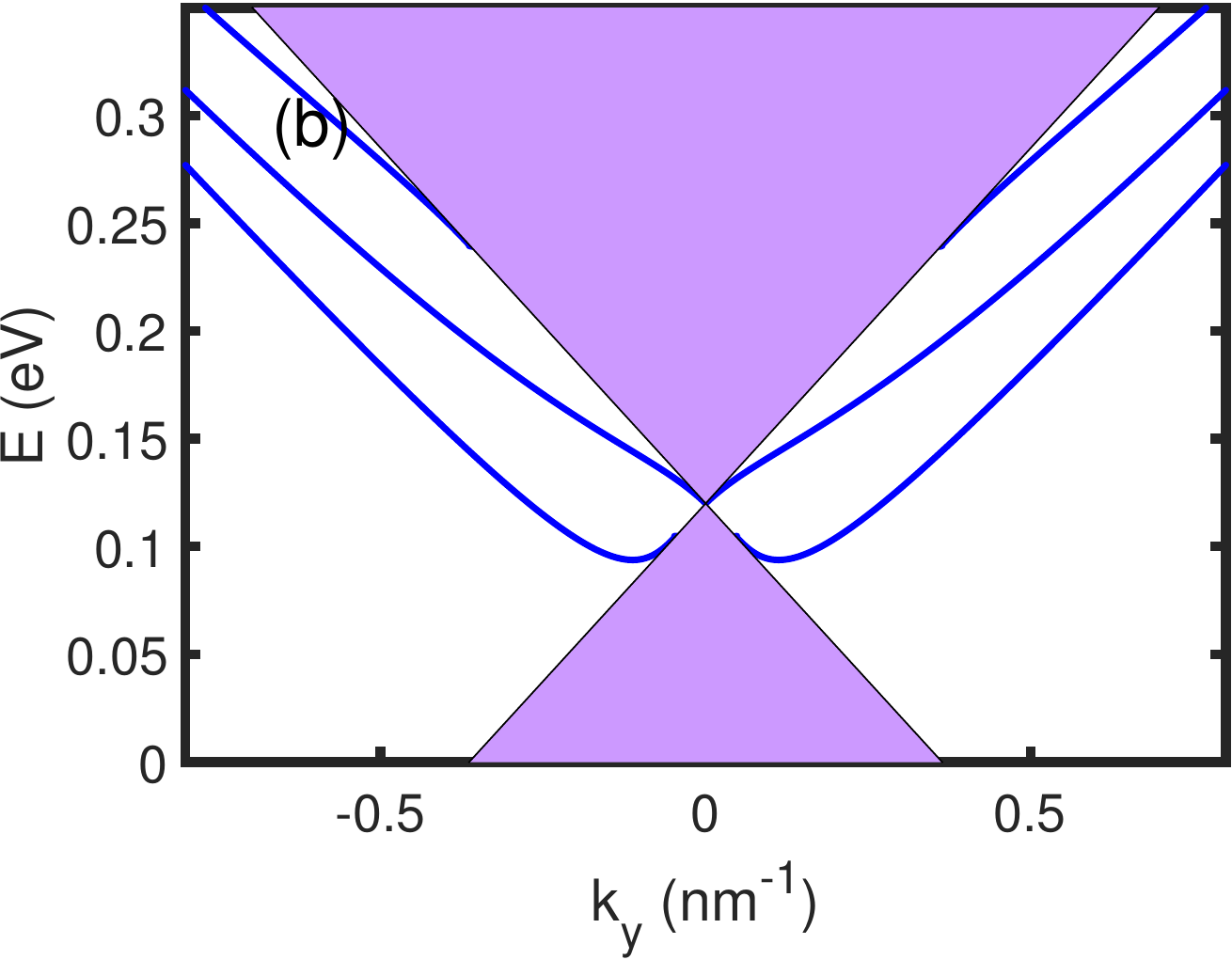}\\
 \includegraphics[width=0.2\textwidth]{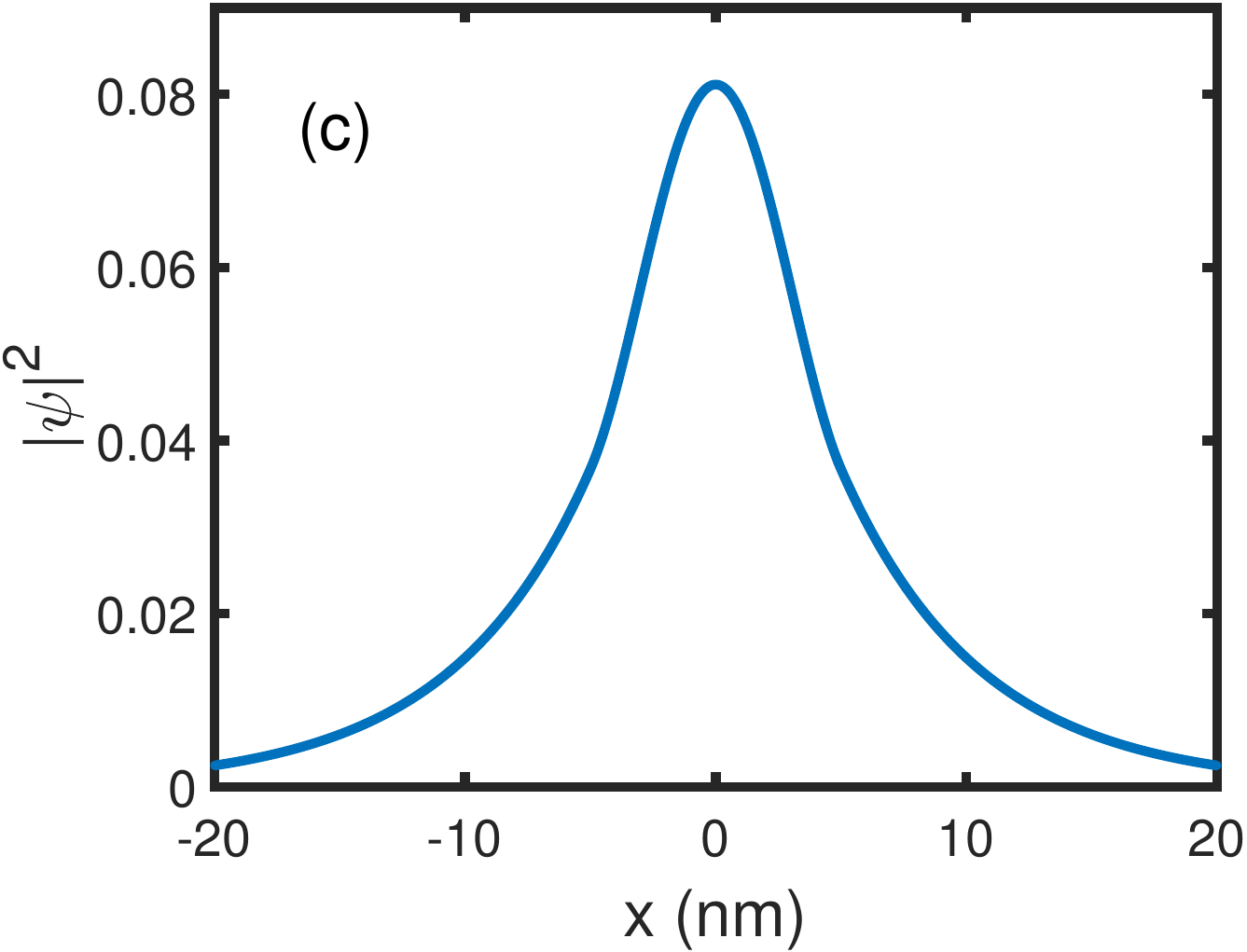}
 \includegraphics[width=0.2\textwidth]{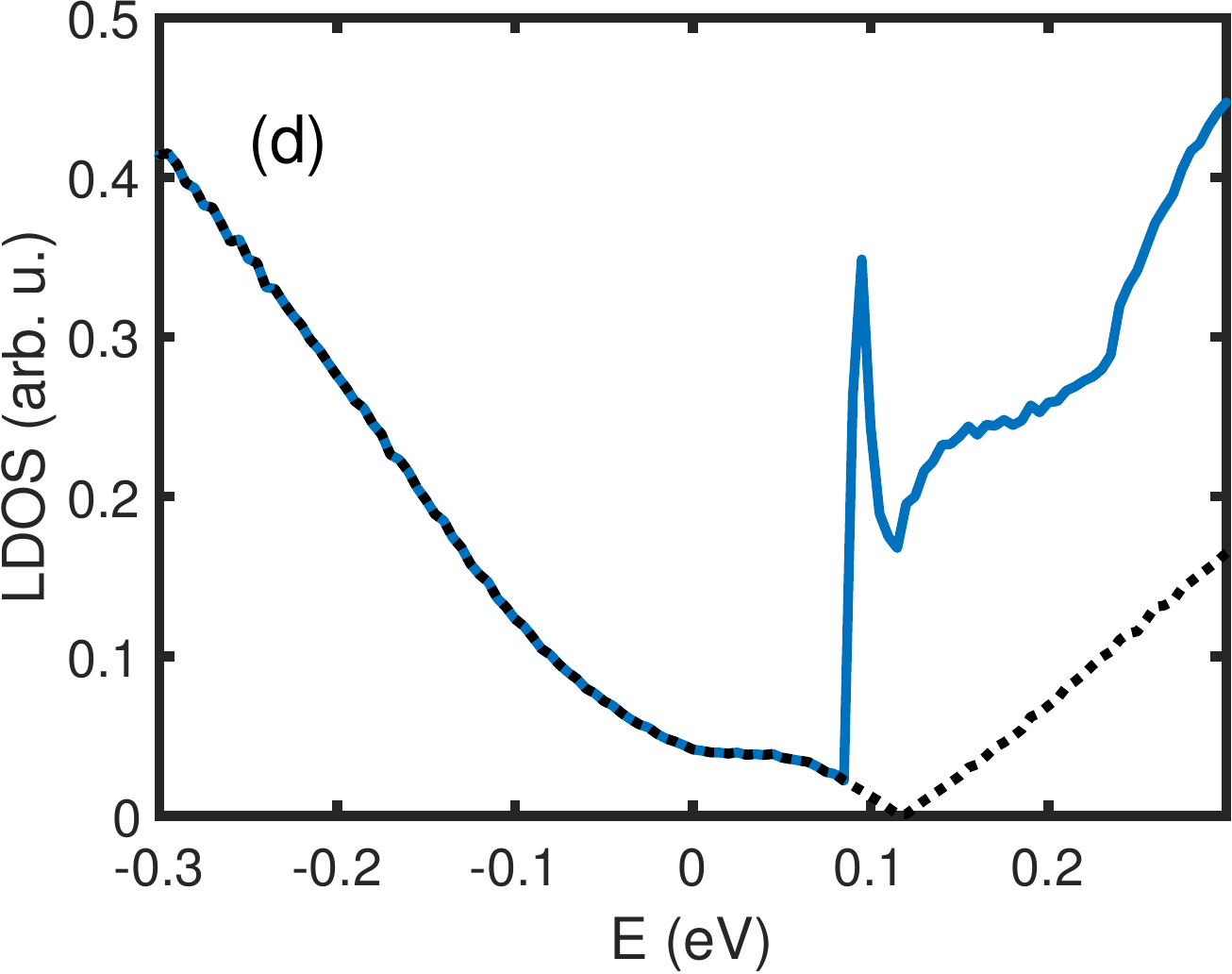}
 \caption{(a) Model square potential well. (b) Dispersion of the bound states in the 
square potential well. (c) A wave function of a bound state from the lowest 
branch in the well. (d)
LDOS in the middle of the rectangular potential well. Dotted line shows the 
contribution of the delocalized states only.}
 \label{SquarePotential}
 \end{figure}

Let us briefly remind the 
specifics of bound states formation in a 
massless Dirac system using an exactly solvable model of a rectangular potential 
well. Following  \cite{Pereira2006,Tudorovskiy2007},
we are considering a system with a Hamiltonian 
\begin{equation}
H=A\boldsymbol{\sigma{}k}+U(x),  
\label{eq:ham}
\end{equation}
where $U$ is a 1D rectangular  potential well running along 
the $y$ axis (Fig.~\ref{SquarePotential}(a))
$$
U(x,y)=
 \left\{
 \begin{aligned}
  &U, x<-l\\
  &0,-l\le x< l\\
  &U,l\le x.
 \end{aligned}
 \right.
$$
Since $H$ is
invariant under translations along the $y$ axis, $k_y$ is a good quantum 
number.  
 Inside the well the 
wave function $\psi$ is a 
combination of $\exp({\pm{}ik_xx+ik_yy})$ and outside the well 
$\psi\propto\exp(-Qx|x|+ik_yy)$.
For $k_x$  and $Q_x>0$ we have $E^2=A^2k_x^2+A^2k_y^2$ 
 and $(E-U)^2=-A^2Q_x^2+A^2k_y^2$ correspondingly.
The continuity condition for the wave function at $x=\pm l$ leads to an 
equation for  $k_{x}$ of the states localized in the $x$ direction in the 
quantum well
\begin{equation}
(EU-A^2k_x^2)\sin{2k_xl}+A^2Q_xk_x\cos{2k_xl}=0.
\label{eq:disp}
\end{equation}
For each value of $k_y$ equation (\ref{eq:disp}) has solutions 
$k_{xn},n=1,2,...$, that give us  branches of the bound 1D states in the 
potential well. 
Corresponding energy dispersions are given by 
$E_n(k_y)=A\sqrt{\left(k_{xn}\right)^2+k_y^2}$.

The dispersions of three lowest branches of these states calculated using 
Eq.~(\ref{eq:disp}) for 
$U=0.12$~eV, $l=5$~nm are shown in Fig.~\ref{SquarePotential}(b). The shaded 
area 
represents the continuum of the 2D states forming the Dirac cone. Attached to it 
are the branches of the bound 1D states. The lowest branch has a minimum, that 
gives rise to a pronounced peak in the density of states. The second branch is 
attached to the Dirac point. For a square potential such a branch exists 
independently of the potential strength, for a sufficiently weak potential it is 
the lowest branch. This may be also the case for other potential shapes as is 
claimed in \cite{Xu2018}.

The calculated local density of states (LDOS) in the center of the potential well is 
shown in Fig.~\ref{SquarePotential}(d) (solid line). It deviates considerably 
from the V-shape of the unperturbed LDOS. The minimum flattens out and a 
maximum appears. The contribution of  delocalized states to the LDOS  is 
shown in Fig.~\ref{SquarePotential}(d) by a dashed line. We see that the sharp 
feature is due to the local density of the bound states. The probability density 
$|\psi|^2$ of a bound state from the lowest branch is shown in 
Fig.~\ref{SquarePotential}(c).

Thus, the signatures of the bound states in the LDOS are disappearance and 
flattening of the sharp
V-shaped minimum, representing the Dirac cone apex, and formation of a single or 
multiple peaks or step-like features (see also \cite{Yampolskii2008}). STM can 
be used to search for such 
features in the tunneling spectra. 
 1D potential wells similar to the one discussed above arise due to the 
band bending in the vicinity of extended surface defects of a topological 
insulator, {\it e.~g.} Bi$_2$Se$_3$.
Below we report our observations of bound states in  
such potential wells~\cite{Fedotov2017}.

For the experimental search for the bound states we performed spatially 
resolved scanning tunneling microscopy and spectroscopy measurements on the 
surface of Bi$_2$Se$_3$ samples cleaved {\it in situ}. All the measurements were 
done at liquid helium temperature in the UHV conditions (typical base vacuum 
$2\times 10^{-11}$~Torr). Pt-Rh tips were used, their quality was checked on Au 
foil. If needed, we performed a tip recovery procedure which included briefly 
dipping the tip into the Au foil followed by the  tip control procedure.
The $dI/dV$ curves of the tunneling junction (tunneling spectra) were obtained 
by numerically differentiating measured $I(V)$ curves. 
To account for the band bending and extract information about the local 
potential we use the normalization method described in Ref.~\cite{Fedotov2017}.
The local potential is obtained as the overall shift of the normalized $dI/dV$ 
curve.

\begin{figure}
\includegraphics[width=0.45\textwidth]{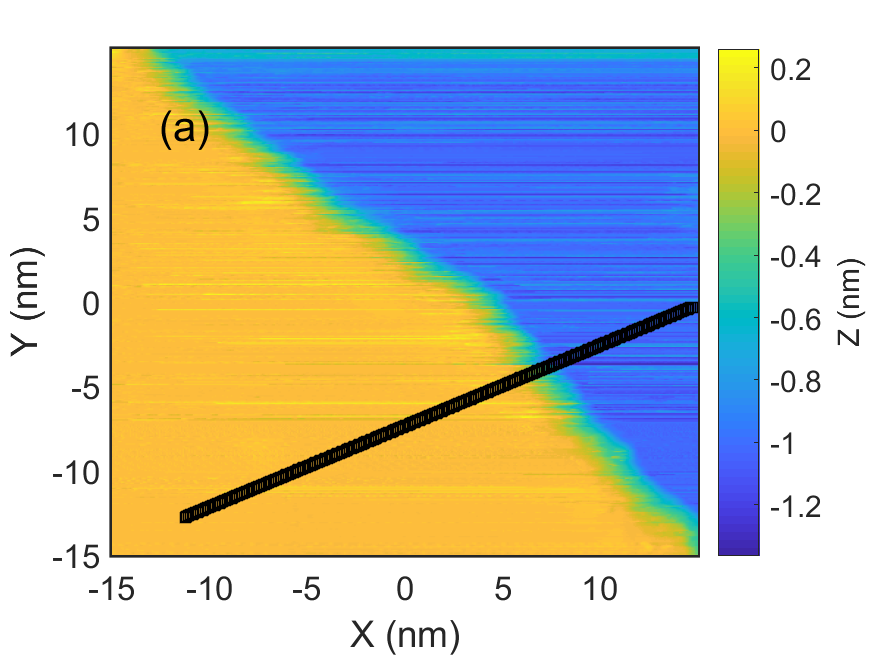}
\includegraphics[width=0.45\textwidth]{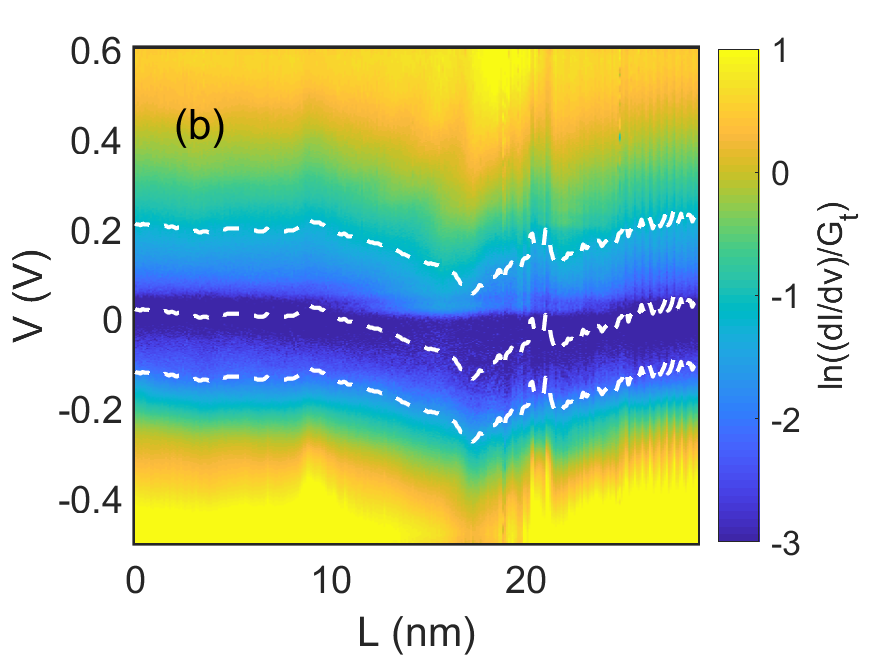}
\includegraphics[width=0.45\textwidth]{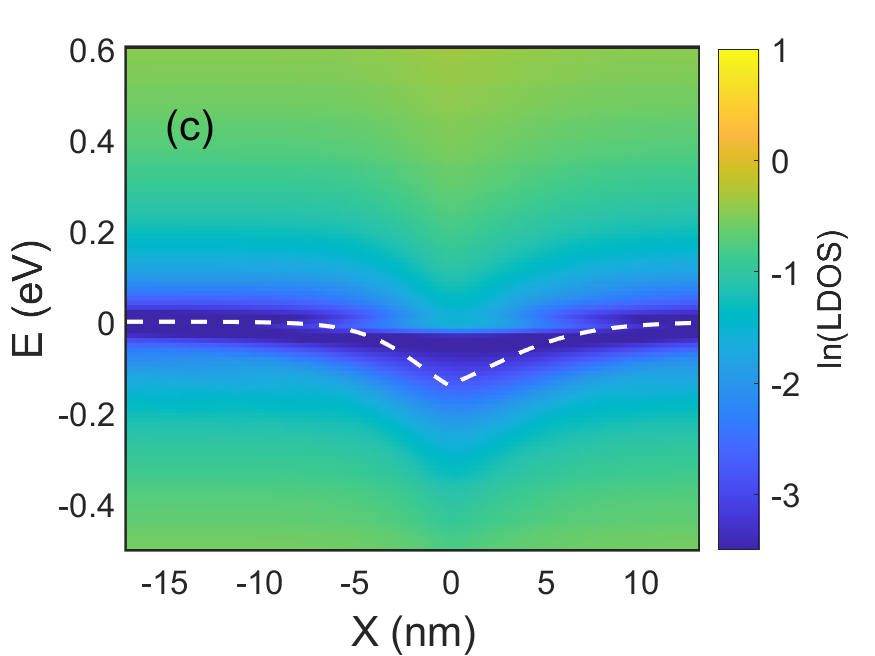}
\caption{(a) STM image of a step on the Bi$_2$Se$_3$ surface. Black squares 
represent the points where I(V) curves were taken. (b) Normalized differential 
tunneling conductance from 300
I-V curves collected along the line across 
the step on the Bi$_2$Se$_3$ surface. White dashed lines 
represent approximate positions of the Dirac point and bulk band edges.  (c) 
LDOS of the surface states
 obtained from numerical calculations for the potential (white 
dashed line) approximating the potential landscape near the step on the 
Bi$_2$Se$_3$ surface. The calculation results are smeared out using a Gaussian function with $\sigma=10$~meV. Note that a contribution of the bulk states is present in (b) but not in (c). Set point $V=0.6$~V, $I=100$~pA. $T=5$~K}
\label{stepPic}
\end{figure}

Bi$_2$Se$_3$ is a layered compound that consists of  quintuple layers 
(QL) Se-Bi-Se-Bi-Se bound one with another  by  van der Waals interaction. 
When Bi$_2$Se$_3$ is cleaved, high steps ($\gtrsim1$ nm) are formed if one or 
more quintuple layers are torn. An STM image of such a step is shown in 
Fig.~\ref{stepPic}(a).
The step height $\sim 1$ nm corresponds to 1 QL.
In the vicinity of these steps on Bi$_2$Se$_3$ surface on a $\sim10$~nm scale a 
$100-200$~meV shift of the chemical 
potential  occurs~\cite{Dmitriev,Fedotov2017}, thus forming a potential well for the Dirac electrons.

\begin{figure}[hb]
\includegraphics[width=0.2\textwidth]{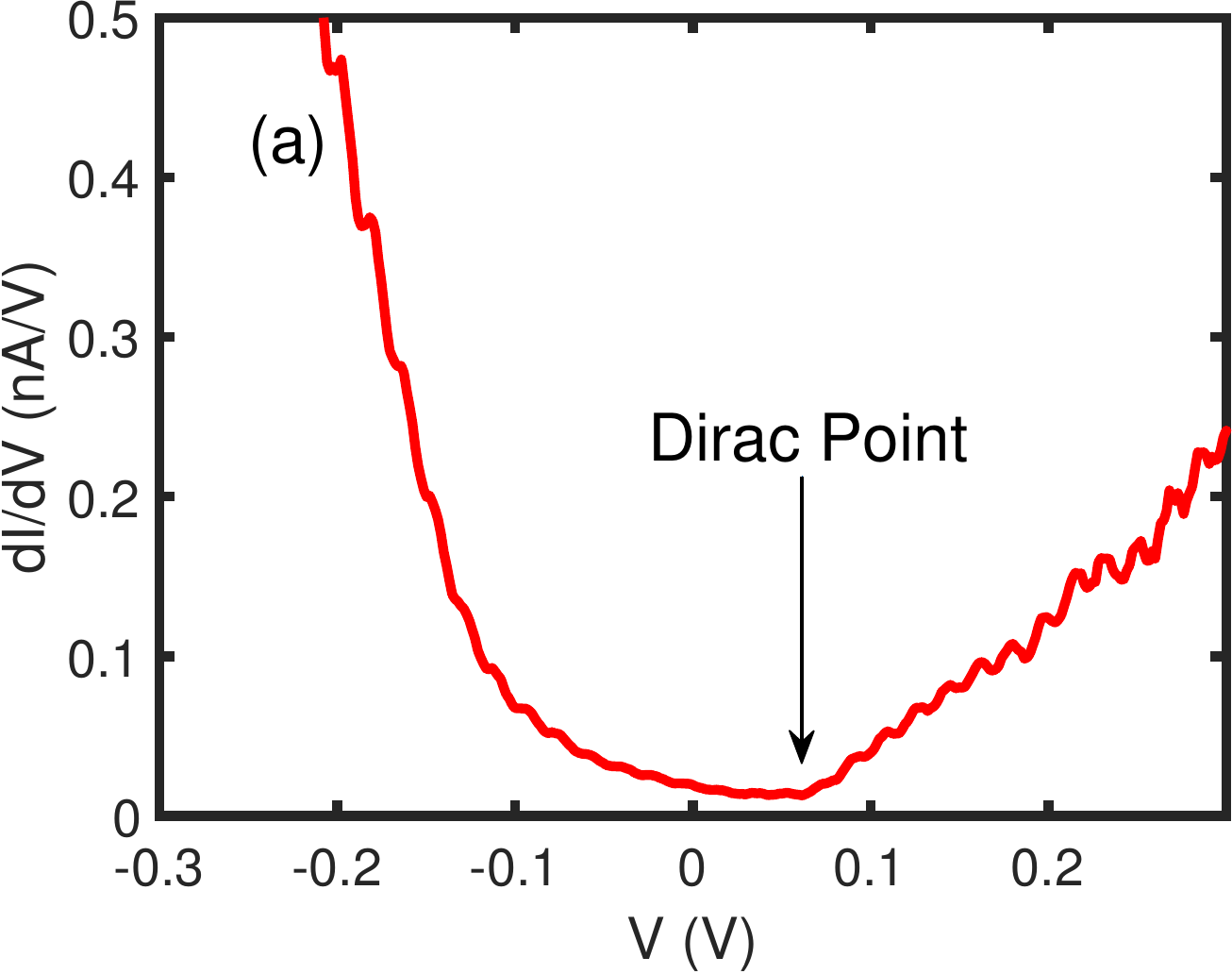}
\includegraphics[width=0.2\textwidth]{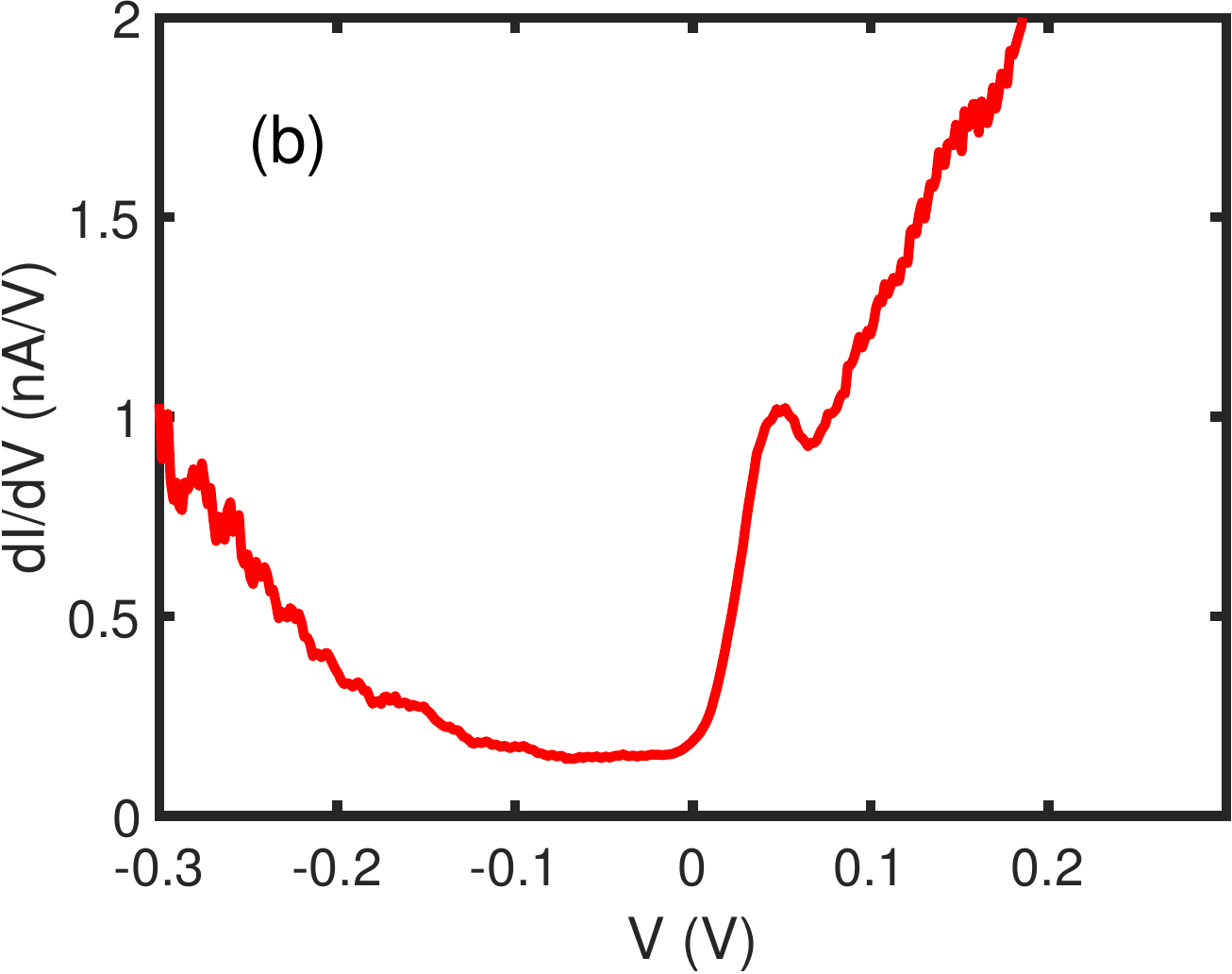}
\caption{(a) A typical $dI/dV$ curve on the surface of the topological insulator 
Bi$_2$Se$_3$ away from defects. (b) A $dI/dV$ curve on the surface 
of the topological insulator 
Bi$_2$Se$_3$ near a surface step. Set point $V=-0.3$~V, $I=100$~pA. $T=5$~K}
\label{vah}
\end{figure}

A typical differential tunneling conductance  ($dI/dV$) curve taken far from 
any defects is presented in Fig.~\ref{vah}(a). 
As the Dirac point of the Bi$_2$Se$_3$ surface states lies within its bulk band 
gap, it corresponds to the V-shaped minimum of the $dI/dV$ curve 
(shown with an arrow). A differential tunneling conductance curve taken on a 
step 
is shown in Fig.~\ref{vah}(b). Apart from an overall shift in 
voltage, 
corresponding to the local potential,
it shows significant change in shape in comparison with the spectrum away from 
defects (Fig.~\ref{vah}(a)). Specifically, the V-shaped minimum corresponding 
to the Dirac point flattens out and a sharp rise with a maximum appears at the 
side of the flattened region of the curve. From the comparison with the model 
predictions it is evident, that these changes of LDOS are in agreement with the 
expected effect of a potential well. In particular the sharp feature corresponds 
to bound states formation.

A spatially resolved STS map taken along a line (black squares in 
Fig.~\ref{stepPic}(a)) across the step in Fig.~\ref{stepPic}(a) is shown in 
Fig.~\ref{stepPic}(b). 
Approximate positions of the Dirac point and bulk band edges (depicted by white 
dashed lines in Fig.~\ref{stepPic}(b)) are determined as in 
Ref.~\cite{Fedotov2017}.
A 0.15~V deep and $\sim15$~nm wide potential well forms due to the band 
bending  in the 
vicinity of the step (which is located at $L\approx17$ nm). 
A horizontal feature of the normalized $dI/dV$ appears in the potential well 
region at $V\approx 0.02$~eV. This feature in the STS map corresponds 
to a maximum of the differential tunneling conductance, such as the one in 
Fig.~\ref{vah}(b). This maximum is a feature of  the LDOS of the sample and not of the tip as it is absent on both sides of the step. Moreover, it does not correspond to a dangling-bond state or a state of an atom adsorbed on the step edge because such states do not spread over distances of $\sim10$~nm. We observe this
behavior in different points along a step and in  multiple samples. For instance, Figs.~\ref{vah} and~\ref{stepPic} were obtained on different samples.
 We argue that this horizontal feature is 
evidence of formation of bound states in a system of massless 2D electrons, 
namely the topologically protected surface states of a topological insulator.  

To justify our interpretation we compare the experimental spatially 
resolved  STS data in the potential well near the surface step in Fig.~\ref{stepPic} with the spatial distributions of numerically 
calculated  local density of states of 2D massless Dirac electrons in a one-dimensional potential $U(x)$ of the same shape. We 
perform our calculations based on  the model 
Dirac Hamiltonian Eq~(\ref{eq:ham}). 
 The potential $U$ is assumed to be constant along the $y$ 
axis in our approximation, so that the 
wave function $\Psi(x,y)=\psi(x)e^{ik_yy}$ and the 2D Dirac equation is reduced 
to a 1D equation for each value of $k_y$.
We numerically solve the corresponding equation by a symmetric finite difference method with periodical boundary conditions. 
Grid discretizations of such equations produce spurious solutions, a problem 
known as fermion doubling. One of the ways to avoid the fermion doubling is to 
add a Wilson mass term $wk^2\sigma_z$ \cite{doubling}. This is the method we 
use in this work.

 The numerically calculated spatial distribution of 
the 
local density of states in the quantum well, corresponding to the potential near 
the step in Fig.~\ref{stepPic}(a), is shown in Fig.~\ref{stepPic}(c). The white dashed line depicts the 
potential profile used for the calculations. The results are in reasonable 
qualitative as well as quantitative
agreement with the experimental $dI/dV$ distribution both in energy and in 
space despite the absence of any fitting parameters. Namely,  a sharp feature  appears in the potential well region 
 in the calculations as well as in the experimental results. The energy 
dispersion $E(k_y)$ resulting from the numerical simulation exhibits similar 
features as the one in Fig.~\ref{SquarePotential}(b). Namely, branches of 
bound states arise, attached to the Dirac cone of the 2D delocalized states. 
The local density of these states produces the horizontal feature in the 
spatial distribution of LDOS.  Similar features can be observed in the potential wells near other types of extended defects.

Formation of such bound states (or waveguide 
states)  was discussed theoretically   in \cite{Yokoyama2010, Seshadri2014} in 
the case 
of topological insulators. These two 
papers focus on the branches of the bound states that connect to the Dirac 
point. We find however that  for the typical parameters of the potential wells in our 
case (100~mV, 10~nm) lower lying branches exist, that provide a larger peak-like
contribution to the LDOS.  Note that energy dispersion and properties of such 
states depend on the 
parameters of the potential and may vary {\em e.~g.} with step height or defect 
type.

 The increase of LDOS near a step on the surface of topological insulator Bi$_2$Te$_3$ observed by
Alpichshev {\em et al.} \cite{Alpichshev} is due  to a purely geometrical effect associated with the  presence of an edge connecting two surfaces at an angle \cite{AlpichshevDiss}. Such an effect is also reproduced by numerical simulation \cite{FedotovZZ2018}. Thus, this increase of LDOS does not correspond to a bound state. In our case the bound state forms in a potential well on the surface of Bi$_2$Se$_3$ and is not intrinsically linked to the presence of the step.

 The comparatively large localization length of the states we observe on the upper surface of the step implies that they are not a manifestation of the states forming on the side surface of a step discussed in \cite{Moon}.

In  other papers \cite{Biswas2011, An2012} the surface step 
is modeled as a 
scattering $\delta$-function barrier. We are considering only the 
experimentally observed potential wells formed on both sides of the step. The 
rationale behind this approach is that the topologically protected surface 
states flow around the step. To take the effect of the step into account more
accurately one needs to consider a three-dimensional model.  Note that a barrier potential at the step also results in bound states formation, but branches of  $E(k_y)$ point in the opposite direction. Correspondingly the energy distribution of LDOS is reversed.

The fact that steps are always present at the surface of a topological insulator inevitably leads to the presence of the bound states.  The role of such states is especially important when the Fermi levels is near the Dirac point where the density of states is low. In particular, they may be involved in scattering of surface electrons, which is associated with a change in the spin direction.
So the spin texture associated with such  defects is an interesting question.
Thus the formation of bound states at the extended defects of the surface 
(especially surface steps) of topological insulators may result in additional 
conductivity and scattering channels and has to be taken into account when 
considering prospects of topological insulator based  quantum devices.

In conclusion, we experimentally observe formation of one-dimensional bound states of 
two-dimensional massless Dirac electrons in potential wells  
in the vicinity of  surface steps on the Bi$_2$Se$_3$ topological 
insulator. Numerical simulations support this conclusion and provide a recipe for their identification. The states form 
branches attached to the Dirac cone and can be identified on spatially resolved 
STS maps as sharp horizontal features with characteristic length $\sim 10$~nm. 

{\bf Acknowledgments}
We are grateful to V.A.~Sablikov for valuable discussion.
The work was carried out with financial support of  RSF (grant  16-12-10335).

\end{document}